\documentclass[aps,twocolumn,pre,showpacs,eqsecnum]{revtex4}
\usepackage{graphpap}
\usepackage[dvips]{graphicx}
\usepackage[dvips]{graphics}
\usepackage{color}

\begin{document}

\title{Raman imaging of doping domains in graphene on SiO$_2$}
 \author{C. Stampfer$^1$\footnote{Corresponding author, e-mail: stampfer@phys.ethz.ch}, L. Wirtz$^{2,3}$, A. Jungen$^4$, D. Graf$^1$, F. Molitor$^1$, C. Hierold$^4$, and K. Ensslin$^1$ }
 \affiliation{$^1$Solid State Physics Laboratory, ETH Zurich, 8093 Zurich, Switzerland, \\
$^2$Institute for Electronics, Microelectronics, and Nanotechnology (IEMN), 
CNRS, 59652 Villeneuve d'Ascq Cedex, France, \\ 
$^3$European Theoretical Spectroscopy Facility (ETSF), E-20018 San Sebasti\'an, Spain, \\
$^4$Micro and Nanosystems, ETH Zurich, 8092 Zurich, Switzerland}
\date{ \today}

 \begin{abstract}
We present spatially resolved Raman images of the
G and 2D lines of single-layer graphene flakes. 
The spatial fluctuations of G and 2D lines are correlated
and are thus shown to be affiliated with local 
doping domains.
We investigate the position of the 2D line 
-- the most significant Raman peak to identify single-layer graphene -- 
as a function of charging up to $\left| n \right| \approx 4\times10^{12}~$cm$^{-2}$. 
Contrary to the G line which exhibits a strong and symmetric stiffening
with respect to electron and hole-doping, the 2D line shows a weak
and slightly asymmetric stiffening for low doping.
Additionally, the line width of the 2D line is, in contrast to the G line, doping-independent making this quantity a reliable measure for identifying single-layer graphene.

 \end{abstract}

 \maketitle

\newpage
Graphene has attracted increasing attention over the last few years~\cite{gei07,kat07}.
Its unique electronic properties~\cite{nov05,zha05}, mainly due to the linear energy vs. momentum dispersion
and the electron hole symmetry near the charge neutrality point, makes it an interesting 
nanomaterial for high mobility electronics~\cite{che07,han07}. 
Raman spectroscopy has proven to be a powerful tool to distinguish
single-layer graphene from few-layer graphene and graphite \cite{fer06,gupta,dav07a,
fer07,gra07b}.
The particular electronic structure of graphite and graphene leads to Kohn-anomalies
in the phonon dispersion at the $\Gamma$ and $K$ points~\cite{pis04,pis06}. 
At $\Gamma$, the Born-Oppenheimer
approximation, which is usually employed for the calculation 
of phonon frequencies, is no longer valid because of the Kohn-anomaly~\cite{lazzeri}. 
This leads to a pronounced stiffening of 
the Raman G line upon positive or negative charging (p or n-doping) of 
the graphene sheet
\cite{ando,pis07,yan07}.
In this paper, we investigate how spatially resolved Raman spectroscopy can 
be used to probe doping domains and local charge fluctuations. 
While electron-hole puddles (i.e. local charge fluctuations) have been predicted to be responsible for the 
finite conductance at vanishing
(average) charge carrier density~\cite{hwang} and have recently been
observed using a scanning single electron transistor~\cite{mar07}, the identification of different doping domains might be desirable to investigate novel graphene devices.
Here, we report on Raman measurements on back gate induced charged graphene 
and on Raman imaging of doping fluctuations of isolated graphene flakes.
We focus on the correlation between the shifts of the G line and the 2D line (or D$^*$ line).
The latter one is the most significant Raman peak in single-layer graphene
\cite{fer06,dav07a}. Within the low charging regime (up to $\pm4\times10^{12}~$cm$^{-2}$) 
obtained in our experiments, the 2D line stiffens for both 
electron and hole charging while its line width (in contrast to the G line) is
not affected by charging. A good correlation between the shift
of G and 2D lines is observed. 
However, the spectral resolution and lateral resolution are not sufficient to resolve electron-hole puddles as shown in Ref.~\cite{mar07}. Therefore we refer to charging (i.e. doping) domains rather than to local electron-hole puddles.

We present both Raman images of isolated graphene and Raman measurements on electrically contacted single-layer graphene on 300~nm SiO$_2$, where highly doped Si is used as back gate.
The samples are prepared by micromechanical cleavage~\cite{dav07b} 
and Raman imaging~\cite{sta07a} is used to select single-layer graphene 
flakes~\cite{fer06,dav07a}. By electron-beam lithography we pattern scanning force microscopy
   \begin{figure}[hbt]\centering
\includegraphics[draft=false,keepaspectratio=true,clip,%
                   width=0.85\linewidth]%
                   {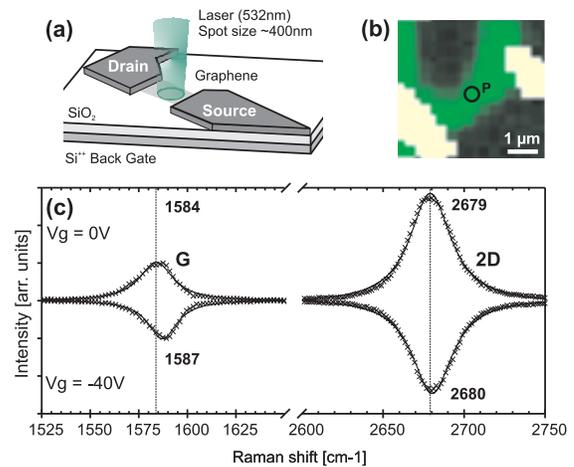}                   
\caption[FIG1]{(color online)
(a) Schematic illustration of a graphene sample with incident laser. (b) Raman image of a contacted single-layer graphene flake, where the integrated 2D line (green; dark) and the inverse Si line (at approx.\ 520~cm$^{-1}$), which is screened by metal electrodes (yellow; bright) are colored. (c) Raman spectra of  the G and 2D line of charge-neutral and charged graphene measured at spot P marked in (b) at T~=~295~K.} 
\label{trdansport}
\end{figure}
pre-mapped electrodes (5 nm Cr/60 nm Au) on the graphene flakes, which 
finally allows to apply a 
back gate voltage $V_g$ between the Si$^{++}$ substrate and graphene.
The experimental setup is shown in Fig.~1(a). The Raman data are recorded by
 using a laser excitation of 532~nm ($E_L$=2.33~eV)
through a single-mode optical fiber, whose spot size is limited by diffraction. 
A long working distance focusing lens with numerical aperture of approx.
0.80 is used to
 obtain a spot size of approx.\ 400~nm. 
We use a laser power below 2~mW such that
 heating effects can be neglected~\cite{jun07a}.
A Raman image of a measured device 
is shown in Fig.~1(b)
and the Raman spectra corresponding to point P (Fig.~1(b)) for a 
charged and charge-neutral case are plotted in Fig.~1(c). The G line shifts approx. by 3~cm$^{-1}$ due to hole charging
  (of $n\approx-4\times10^{12}~$cm$^{-2}$) and a corresponding (hardly visible) small shift in the 2D line is
observed, too. 
\begin{figure}[t]\centering
\includegraphics[draft=false,keepaspectratio=true,clip,%
                   width=0.95\linewidth]%
                   {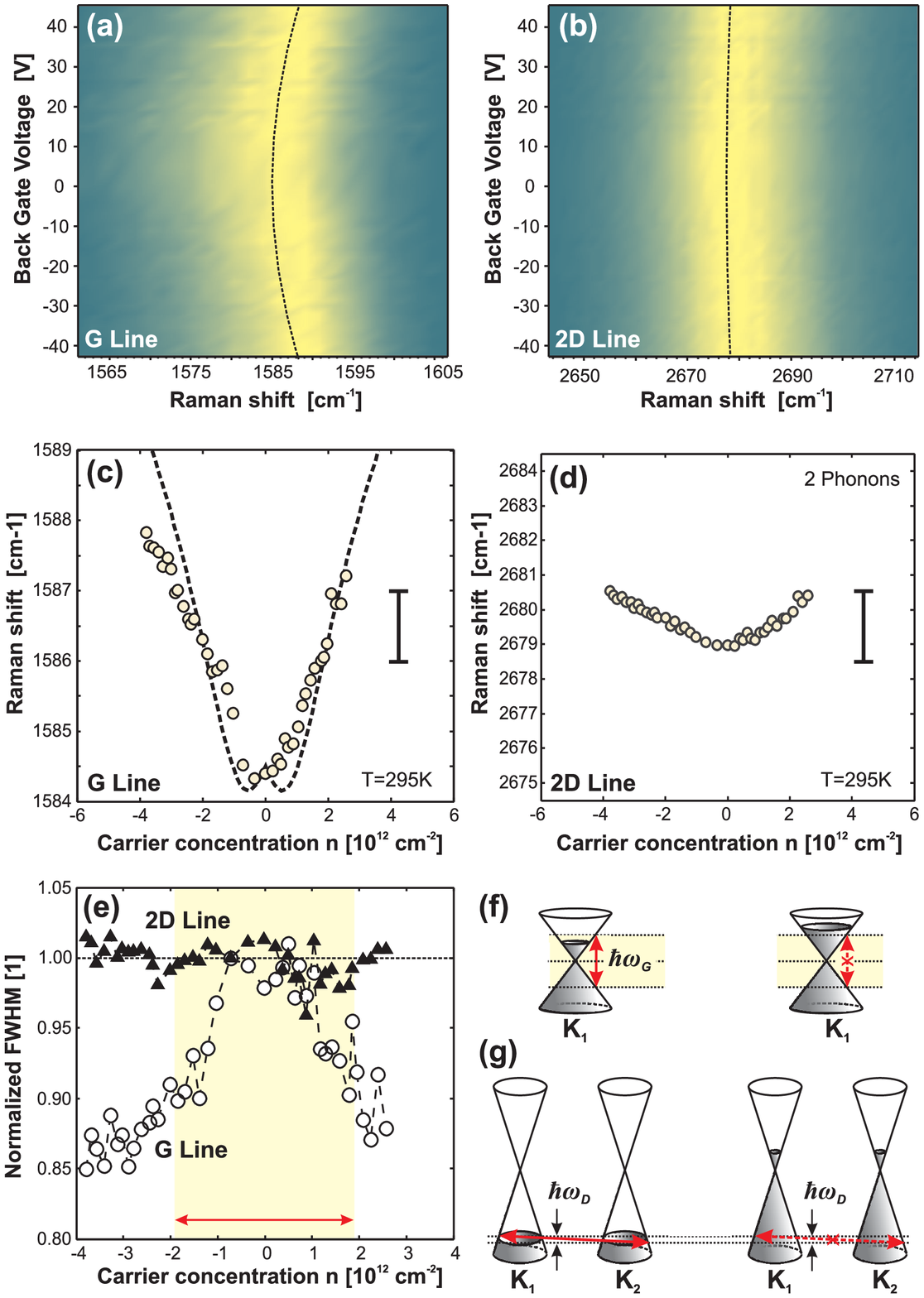}                   
\caption[FIG1]{
(color online) Raman spectroscopy of the G and 2D line of charged graphene.
(a) and (b) show two-dimensional maps of the G and 2D line, respectively as function of 
applied back gate voltages. Both plots are peak height normalized. (c) shows the
G line peak shift [extracted from (a)] as function of the induced carrier concentration.
Dashed lines show theoretical expectations for finite temperature \cite{lazzeri}. (d) same for the 2D peak shift. The same relative frequency-scale as in (c) is used, indicated by the horizontal scale bars (1~cm$^{-1}$/phonon). (e) shows the normalized band width (FWHM) of G and 2D as function of doping. (f) Schematic illustration of the presence (left panel) and absence (right panel) of a $\Gamma$ phonon decay into an electron-hole pair. (g) Presence and absence of D-phonon decay into an electron-hole pair. $K_1$ and $K_2$ denote the two inequivalent $K$-points.
}
\label{expresults}
\end{figure}
Fig.~2 shows Raman data for varying back gate voltages $V_g$ (Figs.~2(a,b)), which by utilizing a simple capacitor model can be substituted by the electron/hole concentration $n=\alpha(V_g-V^D_g)$. Here $\alpha \approx 7.2\times10^{10}$~cm$^{-2}$/V~\cite{nov05} and $V^D_g \approx 2.5$~V marks the charge neutrality point, which has been determined by transport measurements. A typical back gate characteristic of the investigated device is shown in Fig.~4 in Ref.~\cite{mol07b}. The symmetric hole and electron charging-dependent stiffening of the $E_{2g}$ $\Gamma$-phonon was recently explained
as the effect of non-adiabaticity \cite{lazzeri,ando,pis07,yan07},
i.e., the fact that the time-scale of phonon oscillations in graphite
is not long compared to the electron relaxation time (see e.g. Eq.~(6) in Ref.~\cite{pis07}, and dashed line in Fig.~2(c)).
Results of time-dependent perturbation theory were, however,
so far only presented for phonons at the $\Gamma$ point.
For the phonon between $K$ and $M$ which is responsible for the 2D line,
it has been argued \cite{ferrari07a} that non-adiabatic effects are negligible
and the influence of charging can be reproduced by a standard adiabatic
phonon calculation. While non-adiabatic calculations predict a phonon stiffening
for hole-doping and a phonon softening for electron-doping \cite{ferrari07a,
tobepubl}, we observe (see Fig.~2(d)) a slight (asymmetric) stiffening for 
both electron and hole-doping in agreement with the measurements of 
Ref.~\cite{yan07}. 

Another significant difference between the G and 2D lines is their line width 
(FWHM, shown in Fig.~2(e)). 
The G phonon ($q_G \approx 0$) shows a rather strong change as function of 
carrier density~\cite{yan07,pis07} 
which is due to the fact that the Pauli exclusion principle prevents the phonon from decaying into an electron-hole pair 
for $|E_F| > \hbar \omega_G/2$, as illustrated in Fig.~2(f)~\cite{yan07}. 
The decay of the dispersive D phonon with large wave vector $q_D$ 
is unaffected by the Pauli exclusion 
principle for low doping (Fig.~2(g)). It is expected that the 2D line width stays constant up 
to a charging that corresponds to a Fermi level shift as large as the exciting laser energy
$|E_F| \approx E_L$.
This has a practical implication: Since the peak width of the 2D line,
which has been recognized as the most striking feature to distinguish 
single-layer from few-layer graphene~\cite{fer06,gupta,dav07a}, is insensitive 
to doping, it is a reliable - doping-independent - measure for 
identifying single-layer graphene.

\begin{figure}[t]
\centering
\includegraphics[draft=false,keepaspectratio=true,clip,%
                   width=0.99\linewidth]%
                   {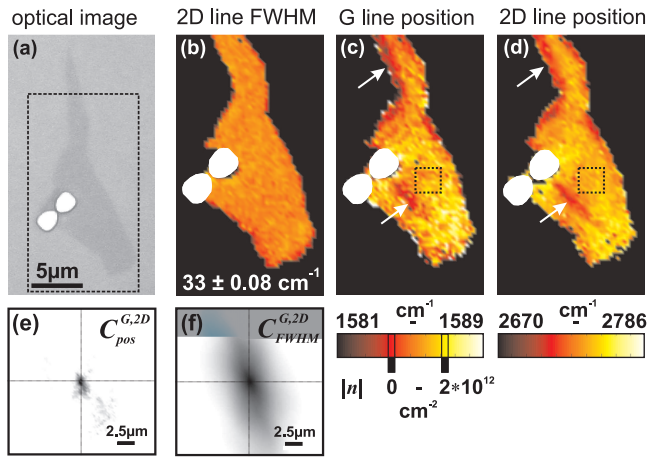}                   
\caption[FIG1]{
(color online) 
Raman images (RIs) of an isolated single-layer graphene flake on SiO$_2$. (a) Green filtered
optical image (the two white dots are metal markers). (b) RI of the FWHM of the 2D line, proving
the single-layer character of the flake. The white dashed line marks the metal marker. (c,d) shows RIs of the G and 2D peak positions, respectively. Most attention is paid to the correlated peak fluctuations, which are due to doping fluctuations. 
The white arrows mark correlated low doping areas, whereas the dotted box highlights a region with low charging fluctuations.
(e) [(f)] Two-dimensional cross-correlation of the fluctuations of the G and 2D line positions [widths]. In both cases the gray scale is adjusted by setting the maximum value to black.
Strong correlation results (e) therefore in a centered peak at (0,0), whereas low concentration leads to a distributed function [e.g. (f)].
} 
\label{setup}
\end{figure}

Spatially resolved Raman spectroscopy (Raman imaging) sensing the doping-dependent G and 2D line shifts provides an interesting tool to
investigate charge fluctuations and doping domains in graphene.
Here we present an example of Raman images (80x45 pixels) of a graphene flake on SiO$_2$ (Fig.~3), where most attention has been paid to the G and 2D peak positions and widths, their fluctuations and cross-correlations. The average G peak position (Fig.~3(c)), $\overline{\omega}_G$, measured on the flake shown in Fig.~3(a), is 1585.4~cm$^{-1}$ and the root mean square (RMS) of the peak fluctuations is~3.3~cm$^{-1}$. This fluctuation also nicely explains the wide spread range of
$\overline{\omega}_G$ reported in the literature~\cite{fer06,gupta,dav07a,
fer07,gra07b}. The 2D line (Fig.~3(d))
is centered around $\overline{\omega}_{2D}=$~2679.4~cm$^{-1}$ and fluctuates with~0.9~cm$^{-1}$. The ratio of the fluctuations of the G and 2D line agrees well with the ratio of the doping-dependent G and 2D stiffening (Figs.~2(c,d))~of~$\approx$ 3.2. 
In Figs.~3(c,d) doped regions on the imaged flake can be observed~\cite{gra07b}. In the upper part we see that towards the edges of the graphene sample charging is suppressed, whereas in the entire inner area significant charging is present. Focusing on a quite uniform area (dotted box in Figs.~3(c,d)) we find that the $\omega_{G}$-fluctuations  are approx.~$0.6$~cm$^{-1}$, which corresponds to $\Delta n \approx 2.4 \times 10^{11}$~cm$^{-2}$.
Please note, that the local charge fluctuations due to electron-hole puddles presented in Ref.~\cite{mar07} are by one order of magnitude smaller in amplitude and have been measured at low temperature and controlled environment.
In addition, our lateral resolution is limited by the laser spot size. 
Figs.~3(c,d), however, nicely illustrate the correlation of G and 2D fluctuations (see e.g. white arrows). This can be also expressed by the two-dimensional cross-correlation function $C^{G,2D}_{\omega}(\Delta \vec{r})=\int \omega_G(\vec{r}+\Delta \vec{r})\; \omega_{2D}(\vec{r}) dx dy$, which exhibits a clear narrow central peak (Fig.~3(e)). In contrast $C^{G,2D}_{FWHM}(\Delta \vec{r})$ does show no measurable correlation of the width of G and 2D (Fig.~3(f)) as expected, since they are uncorrelated quantities.
Indeed the FWHM of the 2D line does not exhibit any doping dependence as shown in Fig.~3(b). 
The line width of the 2D line stays constant at 33$\pm$0.08~cm$^{-1}$ testifying its good quality for identifying single-layer graphene.

In summary, we have presented Raman shifts of the G and 2D line 
for back gate induced charged graphene. We have
discussed the spatial variations in the G and 2D peak positions which 
can be attributed to different doping domains. In the low doping regime no clear 
distinction between electron and hole-doping can be made since for both G and 2D 
line, stiffening is observed. However, absolute doping fluctuations can be estimated. 
This technique is promising to investigate, e.g., electric field distributions 
in (side) gated graphene devices~\cite{mol07}.

{Acknowledgment ---}
The authors wish to thank A.~B\"urli, S. Pisana, A. C. Ferrari,
C. Roman, T.~Helbling, A.~Rubio, M.~Lazzeri, and F. Mauri
for helpful discussions. Support by the ETH FIRST Lab and financial 
support by the TH-18/03-1 grant, 
Swiss National Science Foundation (20021-108059/1) and NCCR nanoscience are gratefully 
acknowledged. L.W. acknowledges support from the French National Research
Agency.

{\em Note added} --- During completion of the manuscript, we became aware of
a quite similar work \cite{ferrari07b} in which doping-dependent fluctuations
and a correlation between the position of G and 2D line are observed
during non-spatially resolved measurements on many different graphene flakes.


\end{document}